\documentclass[10pt,twocolumn]{article}

\usepackage[a4paper,margin=0.68in,columnsep=0.22in]{geometry}
\usepackage{times}
\usepackage{microtype}

\usepackage{amsmath,amssymb,amsfonts}
\usepackage{graphicx}
\usepackage{booktabs}
\usepackage{multirow}
\usepackage{array}
\usepackage{caption}
\usepackage{subcaption}
\usepackage{xcolor}

\usepackage[hidelinks]{hyperref}

\usepackage[compact]{titlesec}
\titlespacing{\section}{0pt}{0.8em}{0.35em}
\titlespacing{\subsection}{0pt}{0.65em}{0.25em}
\titlespacing{\subsubsection}{0pt}{0.5em}{0.2em}

\title{\vspace{-1.0em}
Slice-Profile-Enabled Phase Distribution Graphs for MRI Simulation
\vspace{-0.3em}}

\author{
Snawar Hussain$^{1}$,
        Daniel C. Hoinkiss$^{1}$,
        J\"orn Huber$^{1}$,
        Vicent Kuhlen$^{1}$,
        Matthias G\"unther$^{1, 2}$\\
{\small $^{1}$Fraunhofer Institute for Digital Medicine
MEVIS, Bremen, Germany}\\
{\small $^{2}$University of Bremen,
Bremen, Germany}\\
{\small \texttt{snawar.hussain@mevis.fraunhofer.de}}
}

\date{}

\begin{document}

\twocolumn[
\maketitle

\begin{center}
\begin{minipage}{0.94\textwidth}
\small
\noindent\textbf{Preprint notice.}
This work is intended for submission to a journal for possible publication.
Copyright may be transferred without notice.
This preprint is a working draft that has not yet been peer-reviewed.
The final, definitive version of this manuscript will be published in a peer-reviewed journal.

\vspace{0.7em}
\noindent\textbf{Abstract.}
  MRI simulation often separates two descriptions that are both essential for
  realistic sequence analysis: Bloch dynamics for waveform-resolved
  radiofrequency (RF) excitation, and phase-graph methods for coherence-pathway
  evolution. Extended Phase Graph (EPG) models provide pathway tracking,
  and Phase--Distribution Graphs (PDG) extend this idea to spatially resolved
  $k$-space simulation, but existing PDG formulations rely on
  hard-pulse RF mixing that is \emph{order-local}: the RF pulse mixes
  $F_n^+$, $F_n^-$, and $Z_n$ at a fixed
  coherence order $n$, without coupling different $k_z$ orders. This
  work introduces a unified Bloch-resolved PDG framework for
  slice-profile-aware MRI simulation. A
  scanner-rasterized sequence is partitioned into RF-sensitive Bloch spans and
  non-RF phase-graph spans. For each unique RF span, Bloch dynamics
  are solved on a slice grid to obtain a spatially varying propagator
  $R(z)$. Its Fourier coefficients
  $\mathcal{R}_{\Delta}$, indexed by slice-order offset $\Delta$, are compiled
  into the PDG state graph as sparse cross-order coupling in $k_z$.
  Graph growth is controlled by retaining the dominant Fourier
  coefficients and pruning low-contribution PDG states. This retains
  PDG pathway history and voxel-wise image formation while incorporating shaped
  slice-selective and off-resonant RF behavior. Experiments show
  close agreement with direct one-dimensional Bloch slice-profile evolution
  through repeated excitations while retaining only a few hundred active
  PDG states. Image simulations further illustrate slice-position dependence,
  fat-suppression behavior, measured three-dimensional $B_0$ field maps, and
  comparison with scanner data. The proposed framework
  enables sequence-consistent simulation and signal formation understanding
  in regimes where RF physics, spatial encoding, object
  heterogeneity, and echo-pathway formation interact.

\vspace{0.7em}
\end{minipage}
\end{center}

\vspace{1.0em}
]

\section{Introduction}
Magnetic resonance imaging (MRI) is of the most relevant tools for
medical imaging diagnostic
in clinics. It a noninvasive modality that forms images by
manipulating nuclear spin magnetization with radiofrequency (RF) pulses and
magnetic field gradients and measuring the resulting voltage induced in RF
receive coils. Spatial encoding is achieved by gradient-driven phase accrual,
so the measured signal corresponds to samples of the object’s spatial Fourier
transform ($k$-space).
% Thus, MRI is defined as much by its underlying physics, RF waveforms, gradient
% trajectories, relaxation, off-resonance, and hardware constraints as by its
% reconstruction algorithms.
 Sequence timings
influences image contrast, artifacts, and signal-to-noise ratio. As a result,
fast and accurate \emph{physics-based} simulators are a recurring requirement
in MRI: (i) sequence prototyping and safety-/hardware-constrained design,
(ii) quantitative validation of reconstruction methods under controlled
perturbations, and (iii) generation of realistic training data and
differentiable forward models for learning-based reconstruction and
sequence optimization. Despite this need, MRI simulation involves a
difficult accuracy--cost tradeoff.
Models that resolve the full RF waveform, gradient timing, off-resonance, and
spatial encoding can become computationally expensive, while faster pathway-
based models often rely on simplifying assumptions. The main limitations of the
existing simulation viewpoints are summarized below.

The Bloch equations provide a faithful description of spin dynamics under
time-varying RF and gradients. General-purpose simulators such as JEMRIS and
KomaMRI follow this route by numerically solving Bloch dynamics for flexible
sequence and phantom descriptions, with KomaMRI further improving accessibility
and runtime through Julia-based CPU/GPU parallelization ~\cite{jemris,castillo2023koma}.
These tools are valuable for waveform-resolved simulation, but the computational cost remains
substantial,when simulation must be spatially resolved (many voxels) and
also slice-profile resolved (many spins along $z$) while retaining shaped RF,
overlapping gradients, and off-resonance. This computational burden is a
practical bottleneck in iterative workflows (e.g., design loops, parameter
sweeps, learning, or gradient-based optimization).

Extended Phase Graph (EPG) ~\cite{mathiasandweigel, jhennig} methods offer a
complementary description: instead of simulating
large spin ensembles, they propagate coherence pathways indexed by phase-graph
order. EPG and its variants naturally describe gradient-induced
dephasing by shifts in phase-graph order and capture relaxation with
closed-form updates.
% Two limitations are central for spatially resolved sequence simulation.
% First, classical EPG is not, by itself, a full spatial encoder for 2-D/3-D
% imaging: additional modeling is required to map pathway evolution into
% voxelwise $k$-space samples across an image grid.

In standard EPG, a hard RF pulse mixes the three coefficients
$F_n^+$, $F_n^-$, and $Z_n$ at a fixed coherence order $n$ along slice direction. We refer to this
throughout the paper as \emph{order-local RF mixing}. This approximation is
appropriate when the RF action is spatially uniform. During shaped
slice-selective excitation, however, the RF pulse is applied together with a
slice-select gradient, so the effective RF propagator varies with slice
position. In the phase-graph domain, this spatial dependence produces coupling
between different slice orders. Slice-selective EPG and hybrid
Bloch--EPG formulations have addressed
slice-profile effects within pathway-based signal models. For example, ssEPG
models dynamic slice-profile effects in gradient-crushed transient-state
sequences and was applied to MR fingerprinting, where slice-profile
and $\Delta B_0$ effects
can bias relaxometry estimates~\cite{Ostenson2020ssEPG},
while hybrid Bloch--EPG formulations retain a spatial representation
along the slice direction and combine it with EPG configuration states for
off-resonance, spoiling, and echo-pathway analysis in
quantitative MRI settings.These methods improve
slice-profile-aware signal modeling, but they are not designed as general
image-grid simulators for arbitrary spatial phantoms, scanner-rasterized
sequences, and full \(k\)-space acquisition.

Phase--Distribution Graphs (PDG) address the image-formation side of this
problem by extending phase-graph simulation to spatially resolved \(k\)-space
signals. They provide a graph-based interpretation of phase-graph evolution in
which compact \emph{distributions} are propagated, merged, and stored using
quantized trajectory coordinates. For a given sequence, a PDG can be compiled
on a representative voxel into a sparse transition graph and then replayed
voxelwise to generate spatially resolved \(k\)-space signals efficiently
\cite{pdg}. In the notation used throughout this paper, each distribution
carries a trajectory coordinate
\(\boldsymbol{\kappa}=(k_x,k_y,k_z,t)\), with \(k\) in cycles/m and \(t\) in
seconds.

This machinery is attractive for imaging simulation
%  because the
% sequence-specific
% pathway structure is compiled once and then replayed during voxelwise
% image formation.
but PDG as originally formulated uses order-local hard-pulse RF mixing and
% the RF pulse mixes \(F_n^+\), \(F_n^-\), and \(Z_n\) at fixed coherence order
% \(n\).
does not capture shaped slice-selective RF pulses.

In short, EPG methods have long provided a compact pathway-domain description of
dephasing, RF mixing, and echo formation in repeated-pulse MRI sequences
\cite{mathiasandweigel}. Their strength is that they retain the
coherence-pathway
history without explicitly simulating large ensembles of spins. Bloch
simulation,
in contrast, gives a direct waveform-resolved description of RF excitation and
slice profiles, but becomes expensive when repeated over many voxels, spins,
and sequence repetitions. Recent slice-selective EPG and hybrid Bloch--EPG
formulations have clarified relation between spatially
resolved Bloch evolution and phase-graph coupling while
PDG extends the phase-graph idea toward fast, differentiable, spatially encoded
image simulation \cite{pdg}. The remaining gap is to make the
spatially resolved PDG image-formation framework compatible with
Bloch-resolved slice-selective RF pulses, without losing the sparsity that
makes PDG practical.

The present work addresses this gap by:
\begin{enumerate}
  \item Translating each Bloch-resolved RF span into sparse cross-order
    coupling coefficients that can be compiled into the PDG
    state graph without sacrificing PDG scalability.
  \item This links waveform-resolved RF physics with PDG-based pathway
    tracking and voxelwise image formation. The resulting model keeps the
    echo-pathway history and spatial encoding machinery of PDG, while
    adding slice-profile-aware operators.
  \item supporting slice-profile, off-resonance, species-dependent, and
    3D measured-field effects within the same pathway-based forward model.
\end{enumerate}
The overall workflow is:
(1) a scanner-consistent waveform representation is obtained from the
sequence development framework (Sec.~\ref{sec:seq_partition});
(2) shaped RF spans are isolated and simulated with a Bloch solver over a
$z$ grid to obtain a Cartesian propagator $R(z)$, its channel
basis form $R_{\mathrm{ch}}(z)$, and Fourier coupling coefficients
$\mathcal{R}[\Delta]$ (Sec.~\ref{sec:rf_taps})
(Sec.~\ref{sec:rf_taps});
(3) the taps are applied inside PDG as a sparse set of cross-order offsets
$\Delta\in\mathcal{S}$, and pathway growth is controlled by energy-based
support selection together with PDG-style pruning during graph
compilation (Sec.~\ref{sec:pdg_exec}).
\paragraph*{Key takeaway}
This combination targets the practical regime of interest: shaped
slice-selective RF (accuracy needed) together with spatially resolved imaging
simulation (efficiency needed).

\section{Problem Formulation}
\label{sec:problem_formulation}

\begin{equation}
  \mathbf M(z)=
  \begin{bmatrix}
    M_x(z) & M_y(z) & M_z(z)
  \end{bmatrix}^{\top}
\label{eq:M_def}
\end{equation}
Let ~\eqref{eq:M_def} denote the magnetization along the slice direction with
$z\in[-L/2,L/2]$. During an RF-sensitive interval, the RF waveform and any
overlapping slice-select gradient define a time-dependent effective field.
Neglecting relaxation during this interval, Bloch dynamics induce a rotation
at each slice position,
\begin{equation}
  \mathbf M_{\mathrm{out}}(z)
  =
  R(z)\,
  \mathbf M_{\mathrm{in}}(z),
  \qquad
  R(z)\in \mathrm{SO}(3).
  \label{eq:Rxyz_def}
\end{equation}
Here $R(z)$ acts in Cartesian magnetization coordinates.

EPG and PDG updates are written more naturally in a circular channel basis. We
use the normalized convention as follows:
\begin{equation}
  C_+(z)=\frac{M_x(z)+iM_y(z)}{\sqrt{2}},\qquad
  C_-(z)=\frac{M_x(z)-iM_y(z)}{\sqrt{2}},
  \label{eq:circular_components}
\end{equation}
\begin{equation}
  \mathbf q(z)=
  \begin{bmatrix}
    C_+(z) & C_-(z) & M_z(z)
  \end{bmatrix}^{\top}
  =
  S\,\mathbf M(z),
  \label{eq:q_def}
\end{equation}
\begin{equation}
  S=
  \frac{1}{\sqrt{2}}
  \begin{bmatrix}
    1 & i & 0\\
    1 & -i & 0\\
    0 & 0 & \sqrt{2}
  \end{bmatrix},
  \qquad
  S^{-1}=S^{\mathrm H}.
  \label{eq:S_def}
\end{equation}
The RF propagator in the circular channel basis is therefore
\begin{equation}
  R_{\mathrm{ch}}(z)
  =
  S\,R(z)\,S^{-1}.
  \label{eq:Rch_def}
\end{equation}
% Although $R(z)$ is a real rotation matrix,
% $R_{\mathrm{ch}}(z)\in\mathbb C^{3\times 3}$ because the circular basis is
% complex.
% \subsection{Phase-graph coefficients and slice-order convention}
% \label{sec:pg_order_convention}

Over the finite slice interval of length $L$ along the $z$ axis, the
channel magnetization is
expanded in Fourier orders as:
\begin{equation}
  \mathbf q(z)
  =
  \sum_{n\in\mathbb Z}
  \mathbf Q[n]\,
  e^{i2\pi n z/L},
  \label{eq:q_fourier_series}
\end{equation}
\begin{equation}
  \mathbf Q[n]
  =
  \frac{1}{L}
  \int_{-L/2}^{L/2}
  \mathbf q(z)\,
  e^{-i2\pi n z/L}\,dz.
  \label{eq:Qn_def}
\end{equation}
% \begin{equation}
%   \mathbf Q[n]=
%   \begin{bmatrix}
%     F_n^+ & F_n^- & Z_n
%   \end{bmatrix}^{\top}.
%   \label{eq:Qn_components}
% \end{equation}
Eq. \eqref{eq:Qn_def} contains the transverse and longitudinal
phase-graph coefficients   $[F_n^+  F_n^-  Z_n] ^{\top}$ at coherence
order $n$. With the normalized convention in \eqref{eq:circular_components},
$F_n^\pm$ are the Fourier coefficients of $C_\pm$.
Assuming the slice-select direction be the
$z$ axis, the corresponding physical slice wavenumber is
\begin{equation}
  k_z=n\,d_k,
  \qquad
  d_k=\frac{1}{L}\;\;(\mathrm{cycles/m}).
  \label{eq:dk_def}
\end{equation}
Thus, in the PDG coordinate
$\boldsymbol{\kappa}=(k_x,k_y,k_z,t)$, an RF-induced change of slice order by
$\Delta$ corresponds to
$
k_z \leftarrow k_z+\Delta d_k.
$

\subsection{Order-local hard-pulse RF mixing}
\label{sec:order_local_problem}

In the spatial slice coordinate, the RF action is pointwise:
\begin{equation}
  \mathbf q_{\mathrm{out}}(z)
  =
  R_{\mathrm{ch}}(z)\mathbf q_{\mathrm{in}}(z).
  \label{eq:rf_pointwise}
\end{equation}
In Fourier/phase-graph order, multiplication
by \(R_{\mathrm{ch}}(z)\) becomes convolution over orders.
For a spatially uniform RF propagator, as assumed in the hard-pulse EPG/PDG
update,
\begin{equation}
  R_{\mathrm{ch}}(z)\equiv R_{\mathrm{ch},0}.
  \label{eq:order_local_R}
\end{equation}
The Fourier coefficients of \(R_{\mathrm{ch}}(z)\) are then nonzero only at
\(\Delta=0\) and  gives:
$
\mathcal R[\Delta]
=
R_{\mathrm{ch},0}\,\delta_{\Delta 0}.
\label{eq:hard_pulse_delta}
$
Therefore, the general RF convolution reduces to
\begin{equation}
  \mathbf Q_{\mathrm{out}}[n]
  =
  R_{\mathrm{ch},0}\mathbf Q_{\mathrm{in}}[n].
  \label{eq:order_local_update}
\end{equation}
This is the order-local hard-pulse update: RF mixes the channels
\(F_n^+\), \(F_n^-\), and \(Z_n\) at fixed order \(n\), but does not couple
different slice orders.
\subsection{Slice-selective RF as cross-order coupling}
\label{sec:ssp_crossorder}
\begin{figure*}[t]
  \centering
  \includegraphics[width=0.99\textwidth]{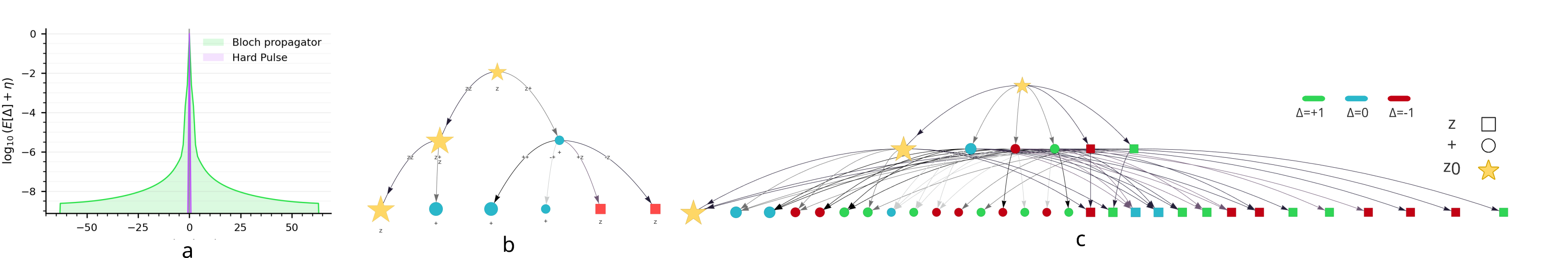}
  \caption{(a) RF tap energy for a spatially uniform RF propagator and a
    slice-selective RF propagator. The spatially uniform case has support only
    at \(\Delta=0\), whereas the slice-selective case has additional support at
    nonzero slice-order offsets because \(R_{\mathrm{ch}}(z)\) varies with
    \(z\).
    Precomputed PDG for the same two FLASH repetitions under
    (b) hard-pulse RF and (c) slice-selective RF. Nodes represent compact
    distributions keyed by $(k_x,k_y,k_z,t)$ and pathway type. With
    $G_z=0$ during RF propagation, only $\Delta=0$ contributes and RF mixing is
    order-local. The slice-selective pulse yields nonzero $\Delta$ taps, spawns
    additional $k_z$ orders, and increases graph branching. For visual clarity,
  only a subset of retained RF offsets is shown.}
  \label{fig:pdg_illustration}
\end{figure*}

Slice-selective excitation has a natural spatial-frequency interpretation
\cite{Pauly1989ExcitationKSpace}. In the present formulation, this appears
through the position-dependent RF propagator along the slice direction,
\begin{equation}
  R_{\mathrm{ch}}(z)\not\equiv R_{\mathrm{ch},0}.
\end{equation}
The RF action is still pointwise in the slice coordinate,
\begin{equation}
  \mathbf q_{\mathrm{out}}(z)
  =
  R_{\mathrm{ch}}(z)\,
  \mathbf q_{\mathrm{in}}(z),
  \label{eq:qout_pointwise}
\end{equation}
Define the Fourier coefficients of the RF propagator as
\begin{equation}
  \mathcal R[\Delta]
  =
  \frac{1}{L}
  \int_{-L/2}^{L/2}
  R_{\mathrm{ch}}(z)\,
  e^{-i2\pi \Delta z/L}\,dz,
  \qquad
  \Delta\in\mathbb Z ,
  \label{eq:Rtap_def}
\end{equation}
where $\Delta$ is an integer slice-order offset. Substituting the Fourier
series of $\mathbf q(z)$ gives
\begin{equation}
  \mathbf Q_{\mathrm{out}}[n]
  =
  \sum_{\Delta\in\mathbb Z}
  \mathcal R[\Delta]\,
  \mathbf Q_{\mathrm{in}}[n-\Delta].
  \label{eq:rf_conv_gather}
\end{equation}
If $R_{\mathrm{ch}}(z)$ is spatially uniform, then
$\mathcal R[\Delta]=0$ for all $\Delta\neq 0$, and
\eqref{eq:rf_conv_gather} reduces to the order-local hard-pulse update in
\eqref{eq:order_local_update}. For slice-selective RF, nonzero coefficients
$\mathcal R[\Delta]$ appear at $\Delta\neq 0$, producing cross-order
coupling. Fig.~\ref{fig:pdg_illustration}-a illustrates this effect by
comparing the RF-tap spectra of a spatially uniform RF pulse and a
slice-selective RF pulse: the former is concentrated at $\Delta=0$, whereas the
latter has additional support at nonzero slice-order offsets
responsible for creating additional pathways.

\section{Methods}

We aim to simulate arbitrary scanner-ready MRI pulse sequences by combining
waveform-resolved RF propagation with sparse PDG execution. To
achieve this, each sequence
is partitioned into RF-active spans, for which slice-resolved
propagators are computed once per
unique span and per $B_0/B_1$/species bin, and \mbox{non-RF} spans;
these are advanced
by PDG operators. The RF propagators are then converted into sparse
cross-order coupling taps in $k_z$, injected into the PDG prepass as RF spawn
operators, and finally evaluated over the image grid using a finite-voxel
slice-collapse signal model.

\subsection{Sequence interpretation and span partitioning}
\label{sec:seq_partition}

gammaSTAR~\cite{Konstandin2025GammaSTAR} exports a pulse sequence \(s\) as a
hardware-valid, rasterized waveform description
\begin{equation}
  \mathcal{W}(s)=
  \{B_1(t_n),\,\mathbf{G}(t_n),\,\mathrm{ADC}(t_n),\,\Delta t\}_{n=0}^{N-1},
  \label{eq:waveforms}
\end{equation}
where \(B_1(t_n)\in\mathbb C\) is the complex RF waveform,
\(\mathbf G(t_n)=[G_x(t_n),G_y(t_n),G_z(t_n)]^\top\) are the gradient
waveforms, \(\mathrm{ADC}(t_n)\) marks sampling instants, and \(\Delta t\) is
the hardware raster. Since the waveform is produced by the scanner-side
sequence framework, RF/gradient limits, slew limits, and timing constraints are
satisfied before simulation.

For each repetition \(r\), the interpreter partitions the rasterized block
\(\mathcal W^{(r)}\) into RF-active spans and non-RF phase-graph spans. The
\(j\)th RF span is the set of raster samples
\begin{equation}
  \mathcal I_{\mathrm B}^{(r,j)}
  =
  \{t_{r,j,\ell}\}_{\ell=0}^{N_{r,j}-1},
  \qquad
  t_{r,j,\ell}=t_{r,j,0}+\ell\Delta t ,
  \label{eq:rf_span_samples}
\end{equation}
with \(N_{r,j}\) samples. On these spans, the original
\(B_1(t_{r,j,\ell})\) and \(\mathbf G(t_{r,j,\ell})\) samples are retained and
passed to the Bloch RF solver, preserving shaped RF waveforms and overlapping
slice-select gradients to produce reusable slice-resolved propagators. The remaining intervals are represented by PDG event
operators parameterized by gradient moments, durations, relaxation/diffusion
factors, and ADC markers for PDG compilation and replay stream.

\subsection{RF-span propagator bank}
\label{sec:rf_bank}
For each RF span, a slice-dependent propagator is computed on a
one-dimensional grid
along the slice-select direction, \[ z_\ell\in[-L/2,L/2],\qquad
L=T_{\mathrm{sl}}+2p_z,\qquad \Delta z=L/N_z . \] The padding \(p_z\)
reduces boundary artifacts and captures side lobes of the slice
profile. The same domain length defines the slice-order spacing
\(d_k=1/L\) used later for RF coupling. During the RF span,
relaxation is handled by operator splitting: the RF interval is
modeled as a rotation, while relaxation is applied in the subsequent
phase-graph/PDG evolution. For each slice position \(z_\ell\), the RF
dynamics are governed by the Bloch equation without
relaxation~\cite{bloch1946nuclear},
\begin{equation} \frac{d\mathbf M(t,z_\ell)}{dt} =
  \gamma_{\mathrm{rad}}\, \mathbf M(t,z_\ell)\times\mathbf
  B_{\mathrm{eff}}(t,z_\ell), \label{eq:bloch_norelax}
\end{equation} where \(\gamma_{\mathrm{rad}}\) is the gyromagnetic
ratio in rad/s/T. The effective field \(\mathbf B_{\mathrm{eff}}\)
contains the transverse RF field, the slice-select gradient
contribution, transmit-frequency offset, static off-resonance,
optional species-dependent frequency offset, and transmit-field
scaling. For a fixed RF span and parameter bin, this defines the
Cartesian propagator
\begin{equation} \mathbf M_{\mathrm{out}}(z_\ell) =
  R(z_\ell)\,\mathbf M_{\mathrm{in}}(z_\ell), \qquad
  R(z_\ell)\in\mathrm{SO}(3). \label{eq:Rxyz_rf_solve}
\end{equation} The propagator is advanced either by a Cartesian
\(\mathrm{SO}(3)\) exponential update or by an equivalent
spinor/\(\mathrm{SU}(2)\) update followed by conversion to Cartesian
rotation form~\cite{Jaynes1955Matrix}. Each finite RF span is
inserted into the PDG graph as a single center-referenced,
slice-dependent RF transition, analogous to an effective RF matrix at
a pulse focus point~\cite{Guenthner2021HybridBlochEPG}. The Bloch
solve uses the rasterized RF waveform and overlapping slice-select
gradient, so slice-profile and within-pulse off-resonance effects are
retained in \(R(z)\). The center-reference convention avoids double
counting the slice-gradient phase when the surrounding non-RF
intervals are later handled by PDG gradient, relaxation, and
diffusion updates. The hard-pulse case is recovered when \(R(z)\) is
spatially uniform, giving \(\mathcal R[\Delta]=0\) for
\(\Delta\neq0\). 

A global RF phase shift does not require a new Bloch solve. Writing
\(B_1(t)=a(t)e^{i\phi_{\mathrm{TX}}(t)}\), a shift \(B_1(t)\mapsto
e^{i\phi_0}B_1(t)\) conjugates the Cartesian propagator by a \(z\)-rotation,
\begin{equation} R(z;e^{i\phi_0}B_1) =
  Z(\phi_0)\,R(z;B_1)\,Z(-\phi_0), \label{eq:rf_phase_conj}
\end{equation} where \(Z(\phi_0)\in\mathrm{SO}(3)\). Propagators are
therefore solved for a canonical phase-stripped waveform and rephased
afterward. Only unique RF spans are propagated. Each span is assigned
a signature based on the raster time step, canonical RF waveform, and
slice-gradient trajectory; a new signature triggers a Bloch solve,
whereas repeated signatures reuse the cached propagator. Spatially
varying system maps are handled by precomputing a propagator bank
over parameter bins \[ \beta=(\Delta f,b,\sigma), \] where \(\Delta
f\) is off-resonance, \(b\) is transmit-field scaling, and \(\sigma\)
is the species label. The stored bank is
\begin{equation} R_{ij\sigma}(z_\ell) \triangleq
  R\!\left(z_\ell;\Delta f_i,b_j,\sigma\right), \qquad
  \ell=0,\ldots,N_z-1 . \label{eq:R_binned}
\end{equation} For voxel-specific values \((\Delta f(\mathbf
r),b(\mathbf r))\), neighboring bin-center rotations are interpolated
on the rotation manifold using the stored unit-quaternion
representation and spherical linear interpolation
\cite{Shoemake1985Quaternion}. The PyTorch implementation batches the
solve over slice positions, species labels, and optional
off-resonance or transmit-scaling bins. The output of this stage is a
reusable bank of center-referenced Cartesian propagators
\(R(z_\ell)\), which are converted to channel-basis Fourier taps in
Sec.~\ref{sec:rf_taps}.

\subsection{From slice propagators to sparse RF coupling taps}
\label{sec:rf_taps}

\paragraph{Channel conversion and Fourier tap extraction.}
The RF coupling taps \(\mathcal{R}[\Delta]\) are obtained
from the Fourier-coefficient definition in \eqref{eq:Rtap_def}. The
block entries are written as
\begin{equation}
  \mathcal R[\Delta]=
  \begin{bmatrix}
    pp_{\Delta} & mp_{\Delta} & zp_{\Delta}\\
    pm_{\Delta} & mm_{\Delta} & zm_{\Delta}\\
    pz_{\Delta} & mz_{\Delta} & zz_{\Delta}
  \end{bmatrix}.
  \label{eq:R_blocks}
\end{equation}
The row of \(\mathcal R[\Delta]\) denotes the output channel, the column denotes
the input channel, and the subscript \(\Delta\) denotes the RF-induced
slice-order offset.
\begin{equation}
  \Delta k_z = \Delta\,d_k,
  \label{eq:delta_kz_def}
\end{equation}

where \(d_k=1/L\) is defined in \eqref{eq:dk_def}. Thus, the RF tap matrices
\(\mathcal{R}[\Delta]\) act directly on the same physical \(k_z\) coordinate
later carried by the PDG state.
Numerically, \(R_{\mathrm{ch}}(z)\) is sampled on the uniform grid
\(z_n=-L/2+n\Delta z\), \(n=0,\dots,N_z-1\), and
\eqref{eq:Rtap_def} is evaluated by the corresponding discrete Fourier
transform,
\begin{equation}
  \mathcal{R}[\Delta]
  \approx
  \frac{1}{N_z}\sum_{n=0}^{N_z-1}
  R_{\mathrm{ch}}(z_n)
  \exp\!\left(-i2\pi\Delta\frac{z_n}{L}\right),
  \Delta\in\mathcal{D},
  \label{eq:R_disc}
\end{equation}
where \(\mathcal{D}\) is the discrete FFT offset set.
% In implementation this is
% computed with a fixed grid-origin convention and FFT shifting.

% \begin{figure}[t]
%   \centering
%   \includegraphics[width=\columnwidth]{figs/tap_energy_spectrum_comparison.png}
%   \caption{RF tap energy for a spatially uniform RF propagator and a
%     slice-selective RF propagator. The spatially uniform case has support only
%     at \(\Delta=0\), whereas the slice-selective case has
% additional support at
%     nonzero slice-order offsets because \(R_{\mathrm{ch}}(z)\) varies with
%   \(z\).}
%   \label{fig:rf_taps}
% \end{figure}

% Dense support in \eqref{eq:rf_conv_gather} leads directly to state growth in the
% PDG prepass because each RF event can spawn many new \(k_z\) orders, as
% depicted in Fig.~\ref{fig:pdg_illustration}. Proposed method
% therefore retains a sparse
% support \(\mathcal{S}\subset\mathbb{Z}\) based on cumulative tap energy. Define
% \begin{equation}
%   E[\Delta]
%   =
%   \sum_{i=1}^{3}\sum_{j=1}^{3}\left|\mathcal{R}[\Delta]_{ij}\right|^2,
%   \qquad
%   p[\Delta]=\frac{E[\Delta]}{\sum_{\Delta'}E[\Delta']}.
%   \label{eq:tapenergy}
% \end{equation}
% Let \(\Delta_{(1)},\Delta_{(2)},\dots\) be the offsets sorted by decreasing
% \(p[\Delta]\). The smallest support size \(K\) is chosen such that
% \begin{equation}
%   \sum_{r=1}^{K}p[\Delta_{(r)}]\ge 1-\epsilon,
%   \label{eq:cumenergy}
% \end{equation}
% and the retained support is
% $
% \mathcal{S}=\{\Delta_{(1)},\dots,\Delta_{(K)}\}.
% $
% The output of this stage is therefore a sparse RF tap set
% \(\{\mathcal{R}[\Delta]\}_{\Delta\in\mathcal{S}}\) per unique RF span and bin.
Dense support in \eqref{eq:rf_conv_gather} increases PDG state growth because
each RF event can spawn multiple \(k_z\) orders. The proposed method therefore
retains only a sparse tap support \(\mathcal S\) based on cumulative tap energy.
For each offset,
\begin{equation}
  E[\Delta]=\|\mathcal R[\Delta]\|_F^2,
  \qquad
  p[\Delta]=\frac{E[\Delta]}{\sum_{\Delta'}E[\Delta']}.
  \label{eq:tapenergy}
\end{equation}
Offsets are sorted by decreasing \(p[\Delta]\), and the smallest set
\(\mathcal S\) satisfying
\begin{equation}
  \sum_{\Delta\in\mathcal S}p[\Delta]\ge 1-\epsilon
  \label{eq:cumenergy}
\end{equation}
is retained. The output is the sparse RF tap set
\(\{\mathcal R[\Delta]\}_{\Delta\in\mathcal S}\) for each unique RF span and
parameter bin.

\subsection{PDG recap and extension to bloch-resolved RF coupling}
\label{sec:pdg_exec}

Phase--Distribution Graphs (PDG) propagate a sparse set of weighted pathway
distributions rather than explicit spin ensembles. Each distribution carries
(i) a complex amplitude \(a\in\mathbb{C}\),
(ii) a component label corresponding to the stored pathway type, and
(iii) a continuous trajectory coordinate
\begin{equation}
  \boldsymbol{\kappa}=(k_x,k_y,k_z,t),
  \label{eq:kappa_def}
\end{equation}
where \(k_x,k_y,k_z\) are in cycles/m and \(t\) is in seconds. Gradient events
translate \(\mathbf{k}=(k_x,k_y,k_z)\) by known increments, whereas free
precession, relaxation, and diffusion update \(a\) and \(t\). To merge
equivalent pathways, \(\boldsymbol{\kappa}\) is quantized into integer keys
and amplitudes are hash-accumulated.
The sparse RF taps from Sec.~\ref{sec:rf_taps} are inserted into the PDG update
as cross-order coupling operators in the slice direction. In gather form, using
the integer slice order \(n\), the RF action is
\begin{equation}
  \mathbf{Q}_{\mathrm{out}}[n]
  =
  \sum_{\Delta\in\mathcal{S}}
  \mathcal{R}[\Delta]\,
  \mathbf{Q}_{\mathrm{in}}[n-\Delta],
  \label{eq:rf_conv_shift}
\end{equation}
where
$
\mathbf{Q}[n]=[F_n^+,F_n^-,Z_n]^\top .
$
In the PDG prepass it is more convenient to evaluate
\eqref{eq:rf_conv_shift} in scatter, or spawn, form. A parent pathway at order
\(n_{\mathrm{in}}\) deposits into
\begin{equation}
  n_{\mathrm{out}} = n_{\mathrm{in}}+\Delta,
  \qquad
  k_{z,\mathrm{out}}=k_{z,\mathrm{in}}+\Delta d_k ,
  \label{eq:spawn_index}
\end{equation}
with \(d_k=1/L\) as defined in \eqref{eq:dk_def}. Equivalently,
\begin{equation}
  \mathbf{Q}_{\mathrm{out}}[n_{\mathrm{in}}+\Delta]
  \;\mathrel{+}=\;
  \mathcal{R}[\Delta]\,
  \mathbf{Q}_{\mathrm{in}}[n_{\mathrm{in}}],
  \qquad
  \Delta\in\mathcal{S}.
  \label{eq:rf_scatter}
\end{equation}
% Equations~\eqref{eq:rf_conv_shift} and \eqref{eq:rf_scatter} describe the same
% convolution operation \eqref{eq:rf_conv_shift} gathers all parents
% contributing
% to a fixed output order, whereas \eqref{eq:rf_scatter} describes the
% implementation as spawning children from each parent state.
PDG uses partition-state method for RF events
\cite{Woessner1961Diffusion,Kaiser1974Diffusion,mathiasandweigel}, an
RF event first splits a parent distribution into
channel-specific child contributions; only afterward are children with identical
keys merged. 

Let
$
  \mathcal C=\{p,m,z\}
$
denote the RF channel labels corresponding to \(F^+\), \(F^-\), and \(Z\),
respectively. A PDG distribution is written as
\[
  d=(\chi,\boldsymbol{\kappa},a),
  \qquad
  \chi\in\mathcal C,
\]
where \(\chi\) is the stored component label, \(\boldsymbol{\kappa}\) is defined
in \eqref{eq:kappa_def}, and \(a\in\mathbb C\) is the complex amplitude.

For a retained RF offset \(\Delta\in\mathcal S\), a parent distribution
\(d=(\chi,\boldsymbol{\kappa},a)\) produces one child contribution for each
output channel \(\chi'\in\mathcal C\):
\begin{equation}
  d_{\chi\rightarrow\chi'}^{(\Delta)}
  =
  \left(
    \chi',
    \boldsymbol{\kappa}+\Delta d_k\,\mathbf e_{k_z},
    a\,c_{\chi\rightarrow\chi'}[\Delta]
  \right),
  \label{eq:rf_partition_spawn}
\end{equation}
where \(\mathbf e_{k_z}=(0,0,1,0)\). The coefficient
\(c_{\chi\rightarrow\chi'}[\Delta]\) denotes the source-to-destination RF
coupling from input channel \(\chi\) to output channel \(\chi'\) at slice-order
offset \(\Delta\). With the channel ordering
$
  \iota(p)=1, \iota(m)=2, \iota(z)=3,
$
this coefficient is
\begin{equation}
  c_{\chi\rightarrow\chi'}[\Delta]
  =
  \mathcal R[\Delta]_{\iota(\chi'),\,\iota(\chi)} .
  \label{eq:coeff_index_map}
\end{equation}

% Thus the row of \(\mathcal R[\Delta]\) gives the output channel and the column
% gives the input channel.
Children with the same output label \(\chi'\) and the same quantized
\(\boldsymbol{\kappa}\) key are then merged by summing their amplitudes. Their
individual parent contributions are retained as ancestor edges in the PDG graph,
which are later used for pruning and latent-signal scoring.

With the source-to-destination convention, the relevant block entries
of \(\mathcal{R}[\Delta]\) are arranged as in \eqref{eq:R_blocks}.
% \(zp_{\Delta}\) denotes a
% \(Z\rightarrow F^+\) contribution, whereas \(pz_{\Delta}\) denotes an
% \(F^+\rightarrow Z\) contribution. For an input order \(n\) and
% retained offset
% \(\Delta\in\mathcal S\),
Writing the RF update in the same, a parent at
\(k_{z,\mathrm{in}}\) contributes separate child amplitudes
\begin{equation}
  \begin{aligned}
    F^+_{\mathrm{out}}[n+\Delta]
    &\mathrel{=}
    pp_{\Delta}\,F^+_{\mathrm{in}}[n],\\
    F^+_{\mathrm{out}}[n+\Delta]
    &\mathrel{=}
    mp_{\Delta}\,F^-_{\mathrm{in}}[n],\\
    F^+_{\mathrm{out}}[n+\Delta]
    &\mathrel{=}
    zp_{\Delta}\,Z_{\mathrm{in}}[n],
  \end{aligned}
  \label{eq:rf_Fp_spawn}
\end{equation}
and the spawned contributions to the stored longitudinal output are
\begin{equation}
  \begin{aligned}
    Z_{\mathrm{out}}[n+\Delta]
    &\mathrel{=}
    pz_{\Delta}\,F^+_{\mathrm{in}}[n],\\
    Z_{\mathrm{out}}[n+\Delta]
    &\mathrel{=}
    mz_{\Delta}\,F^-_{\mathrm{in}}[n],\\
    Z_{\mathrm{out}}[n+\Delta]
    &\mathrel{=}
    zz_{\Delta}\,Z_{\mathrm{in}}[n].
  \end{aligned}
  \label{eq:rf_Z_spawn}
\end{equation}
Each line in \eqref{eq:rf_Fp_spawn}--\eqref{eq:rf_Z_spawn} is a separate
spawned edge before merging. After hash-based merging, these contributions are
algebraically equivalent to the corresponding rows of the compact convolution
in \eqref{eq:rf_conv_shift}. The \(F^-\) output row
\((pm_{\Delta},mm_{\Delta},zm_{\Delta})\) is not stored as a separate
distribution family; it is represented through
the implicit conjugate branch.
For physical magnetization,
$
F^-_{n}=\overline{F^+_{-n}},
\label{eq:Fminus_pair}
$
In trajectory coordinates, this conjugate pathway is
represented by
\begin{equation}
  (k_x,k_y,k_z,t,a)
  \;\mapsto\;
  (-k_x,-k_y,-k_z,-t,\overline{a}).
  \label{eq:pdg_conj_map}
\end{equation}
% Thus, terms in \eqref{eq:rf_Fp_spawn} and \eqref{eq:rf_Z_spawn} involving
% \(F^-_{\mathrm{in}}\) are evaluated by generating the mirrored parent from the
% stored \(F^+\) distribution using \eqref{eq:pdg_conj_map}, applying the
% appropriate \(m\rightarrow p\) or \(m\rightarrow z\) coefficient, and then
% spawning the child with the same additive order shift \(+\Delta\). The RF tap
% offset is therefore not negated; the conjugation is applied to the parent
% coordinate and amplitude before the usual scatter update.

% Therefore, Whenever an \(m\)-type parent is
% needed, it is
% generated on demand from the conjugate relation in \eqref{eq:pdg_conj_map}.
For example, a stored \(p\)-type parent can produce a direct \(p\rightarrow p\)
child and, through its mirrored conjugate parent, an \(m\rightarrow p\) child:
\begin{align}
  d_{p\rightarrow p}^{(\Delta)}
  &=
  \left(
    p,\,
    \boldsymbol{\kappa}+\Delta d_k\,\mathbf e_{k_z},\,
    a\,pp_{\Delta}
  \right),\\
  d_{m\rightarrow p}^{(\Delta)}
  &=
  \left(
    p,\,
    \bar{\boldsymbol{\kappa}}+\Delta d_k\,\mathbf e_{k_z},\,
    \bar a\,mp_{\Delta}
  \right),
\end{align}
where
\(\bar{\boldsymbol{\kappa}}=(-k_x,-k_y,-k_z,-t)\). These are distinct spawned edges before merging. They
are accumulated into the same \(p\)-type output container only because both
produce an \(p\)-type child. 
% If their output keys differ, they remain separate PDG
% nodes; if their output type and quantized \(\boldsymbol{\kappa}\) key are the
% same, their amplitudes are summed and both ancestor edges are retained.
This follows the standard PDG
bookkeeping.

\paragraph{Graph compilation and voxelwise replay.}
Compiling PDG involves constructing a sparse computation graph by
propagating distributions through repetitions and recording their
ancestor relations.
The graph edges encode
(i) gradient translations,
(ii) free-precession, relaxation, signal collection and diffusion
updates over event durations for representative voxels,
and
(iii) RF spawn edges via partition state method induced by the retained offsets
$\Delta\in\mathcal{S}$.
During compilation, $\boldsymbol{\kappa}$ is quantized to integer keys and
merged in hash maps. Execution then replays this fixed sparse transition
structure over the image grid, applying voxel-specific parameters
(e.g. relaxation, diffusion, off-resonance, transmit scaling, and species)
while preserving the same graph topology.

\subsection{Extension to signal equation along the slice direction}
\label{sec:measurement_model}

Proposed method inherit the PDG state representation, graph
compilation, voxelwise replay,
and latent-signal pruning from baseline formulation.\cite{pdg}. The extension
introduced in the proposed method changes the RF transition:
Bloch-resolved RF pulses populate nonzero slice orders \(k_z\).
Consequently, the signal model must evaluate the slice-direction part of the
spatial basis explicitly.
%  In the original hard-pulse setting this term is often
% trivial because the RF update does not generate additional \(k_z\) orders.
Following the PDG notation, a single state of type
\(\hat e\in\{+,-,z\}\) is written as
\begin{equation}
  M^{\hat e}(\mathbf r,\omega)
  =
  F^{\hat e}_{\mathbf k,\tau}\,
  e^{\,i2\pi\mathbf k\cdot\mathbf r}\,
  e^{\,i\tau\omega}\,
  W(\omega,\mathbf r)\,
  V(\mathbf r),
  \label{eq:pdg_general_state}
\end{equation}
where \(F^{\hat e}_{\mathbf k,\tau}\in\mathbb C\) is the state coefficient,
\(\mathbf k=(k_x,k_y,k_z)\) is the spatial modulation carried by the state,
\(\tau\) is the off-resonance dephasing coordinate,
\(W(\omega,\mathbf r)\) is the local off-resonance distribution, and
\(V(\mathbf r)\) is the spatial voxel basis function.
% The \(F^-\) branch is represented implicitly as described in
% Sec.~\ref{sec:pdg_exec}.
For a spatially resolved phantom, the voxel-resolved magnetization is obtained
by summing over voxels \(v\),
\begin{equation}
  M^{\hat e}(\mathbf r,\omega)
  =
  \sum_v\sum_{\mathbf k,\tau}
  \big(F^{\hat e}_{\mathbf k,\tau}\big)_v\,
  e^{\,i2\pi\mathbf k\cdot\mathbf r}\,
  e^{\,i\tau\omega}\,
  W_v(\omega)\,
  V_v(\mathbf r).
  \label{eq:pdg_spatially_resolved}
\end{equation}
This is the original PDG spatially resolved state representation, but with
state support that can now include RF-induced nonzero \(k_z\) orders.
The measured transverse signal is obtained by integrating over space and
off-resonance frequency,
\begin{equation}
  S^+
  =
  \int_V\int_\Omega M^+(\mathbf r,\omega)\,d\omega\,d\mathbf r .
  \label{eq:pdg_signal_general_a}
\end{equation}
\begin{equation}
  \resizebox{\linewidth}{!}{$
    S^+
    =
    \sum_v \sum_{\mathbf k,\tau}
    \big(F^+_{\mathbf k,\tau}\big)_v
    \int_V \int_\Omega
    e^{\,i2\pi\mathbf k\cdot\mathbf r}
    e^{\,i\tau\omega}
    W_v(\omega)
    V_v(\mathbf r)
  \,d\omega\,d\mathbf r$}
  \label{eq:pdg_signal_general}
\end{equation}
% Substituting \eqref{eq:pdg_spatially_resolved} in
% \eqref{eq:pdg_signal_general_a} gives \eqref{eq:pdg_signal_general}
% and using the same Cauchy/Lorentzian model as in the original PDG work, the
% frequency integral yields
% \begin{equation}
%   \int_\Omega e^{\,i\tau\omega}W_v(\omega)\,d\omega
%   =
%   e^{\,i(\Delta\omega_0)_v\tau}\,
%   e^{-|\tau|/T'_{2,v}} .
%   \label{eq:pdg_omega_integral}
% \end{equation}
% \begin{equation}
%   \Phi_v(\mathbf k)
%   =
%   \int_V e^{\,i2\pi\mathbf k\cdot\mathbf r}
%   V_v(\mathbf r)\,d\mathbf r
%   \label{eq:spatial_basis_factor}
% \end{equation}
% \begin{equation}
%   \resizebox{\linewidth}{!}{$
%     S^+
%     =
%     \sum_v \sum_{\mathbf k,\tau}
%     \big(F^+_{\mathbf k,\tau}\big)_v\,
%     e^{\,i(\Delta\omega_0)_v\tau}
%     e^{-|\tau|/T'_{2,v}}\,
%   \Phi_v(\mathbf k).$}
%   \label{eq:pdg_signal_spatial}
% \end{equation}
% The factor \(\Phi_v(\mathbf k)\) is the Fourier response of the voxel basis.
% For a shifted voxel basis \(V_v(\mathbf r)=V_0(\mathbf r-\mathbf r_v)\),
% \begin{equation}
%   \Phi_v(\mathbf k)
%   =
%   e^{\,i2\pi\mathbf k\cdot\mathbf r_v}\,
%   \widehat V_0(\mathbf k).
%   \label{eq:phi_shifted_basis}
% \end{equation}
Substituting \eqref{eq:pdg_spatially_resolved} into
\eqref{eq:pdg_signal_general_a} gives \eqref{eq:pdg_signal_general}. Using the
same Cauchy/Lorentzian model as in the original PDG work, the frequency
integral and the remaining spatial factor are
{\setlength{\jot}{2pt}
  \begin{align}
    \int_\Omega e^{\,i\tau\omega}W_v(\omega)\,d\omega
    &=
    e^{\,i(\Delta\omega_0)_v\tau}\,
    e^{-|\tau|/T'_{2,v}},
    \label{eq:pdg_omega_integral}\\
    \Phi_v(\mathbf k)
    &=
    \int_V
    e^{\,i2\pi\mathbf k\cdot\mathbf r}
    V_v(\mathbf r)\,d\mathbf r .
    \label{eq:spatial_basis_factor}
\end{align}}
Hence,
{\setlength{\jot}{2pt}
  \begin{align}
    S^+
    &=
    \sum_v \sum_{\mathbf k,\tau}
    \big(F^+_{\mathbf k,\tau}\big)_v\,
    e^{\,i(\Delta\omega_0)_v\tau}
    e^{-|\tau|/T'_{2,v}}\,
    \Phi_v(\mathbf k).
    \label{eq:pdg_signal_spatial}
\end{align}}
The factor \(\Phi_v(\mathbf k)\) is the Fourier response of the voxel basis.
For a shifted voxel basis \(V_v(\mathbf r)=V_0(\mathbf r-\mathbf r_v)\),
\begin{equation}
  \Phi_v(\mathbf k)
  =
  e^{\,i2\pi\mathbf k\cdot\mathbf r_v}\,
  \widehat V_0(\mathbf k).
  \label{eq:phi_shifted_basis}
\end{equation}

% This corresponds to the \(e^{i\mathbf k\cdot\mathbf r_v}\Theta(\mathbf k)\)
% term in the original PDG signal equation. Where \(V(\mathbf r)\)
% represents the spatial voxel
% basis. The Fourier transform of this basis gives the voxel-shape response in
% the signal equation. For the commonly used sinc-shaped voxel basis, this
% response corresponds to a box-shaped support in \(k\)-space, i.e., an effective
% Nyquist cutoff determined by the voxel size. This treatment is appropriate when
% large \(\mathbf k\)-values mainly represent intravoxel dephasing generated by
% gradients such as the original PDG hard-pulse RF setting,
% In the present formulation, however, \(k_z\) is not only a
% dephasing coordinate. Nonzero \(k_z\) orders are also generated by the
% slice-selective RF taps \(\mathcal R[\Delta]\), and therefore carry
% slice-profile information. Applying the same hard slice-direction support to
% all \(k_z\) content can then remove physically meaningful RF-induced
% components.
This is the analogue of the
\(e^{i\mathbf k\cdot\mathbf r_v}\Theta(\mathbf k)\) factor in the original PDG
signal equation. Here \(V_v(\mathbf r)\) is the voxel basis, and
\(\Theta(\mathbf k)\) is its Fourier-domain response. For the sinc-shaped voxel
basis used in PDG, this response is a box-shaped \(k\)-space support, i.e., a
Nyquist cutoff determined by the voxel size. This is appropriate when large
\(\mathbf k\)-values mainly represent gradient-induced intravoxel dephasing, as
in the hard-pulse PDG setting. In the present model, however, \(k_z\) also
contains RF-induced slice-profile orders generated by
\(\mathcal R[\Delta]\). Applying the same hard slice-direction cutoff to these
orders can remove physically meaningful RF-induced signal components.

% This effect was observed in species-dependent simulations as shown in
% the Fig \ref{fig:signal_model_comparison}.Water and fat were
% treated as separate
% chemical-shift species, with the fat resonance offset from the water resonance
% as in standard chemical-shift MRI~\cite{Dixon1984WaterFat}.When
% this water--fat
% frequency offset was included during RF propagation, fat-related contributions
% could move into higher nonzero \(k_z\) orders and be attenuated by the hard
% slice-direction cutoff even without a fat-suppression preparation. The
% slice-direction response is therefore evaluated separately through a factor
% \(\Psi_v(k_z)\), either analytically or by explicit \(z\) sampling,
% rather than
% using the same cutoff as the original hard pulse model as described below.
For a two-dimensional phantom with an assumed slice support, write
\(\mathbf r=(\mathbf r_{xy},z)\) and
\(\mathbf k=(\mathbf k_{xy},k_z)\). We factorize the voxel basis as
\begin{equation}
  V_v(\mathbf r)
  =
  V_{v,xy}(\mathbf r_{xy})\,U_v(z).
  \label{eq:voxel_factorization}
\end{equation}
The corresponding spatial response separates into in-plane and slice factors,
{\setlength{\jot}{2pt}
  \begin{align}
    \Phi_v(\mathbf k)
    &=
    \Phi_{v,xy}(\mathbf k_{xy})\,\Psi_v(k_z),
    \label{eq:spatial_factorization}\\
    \Phi_{v,xy}(\mathbf k_{xy})
    &=
    e^{\,i2\pi\mathbf k_{xy}\cdot\mathbf r_{v,xy}}\,
    \Theta_{v,xy}(\mathbf k_{xy}),
    \label{eq:phi_xy_theta}\\
    \Psi_v(k_z)
    &=
    \int e^{\,i2\pi k_z z}\,U_v(z)\,dz .
    \label{eq:slice_collapse_general}
\end{align}}
Here \(\Theta_{v,xy}\) is the in-plane voxel-basis response inherited from the
original PDG measurement model, while \(\Psi_v(k_z)\) is the slice-collapse
factor. For a uniform slice of thickness \(T_{\mathrm{sl}}\) centered at
\(z_v\),
\begin{equation}
  \Psi_v(k_z)
  =
  e^{\,i2\pi k_z z_v}\,
  \mathrm{sinc}(k_z T_{\mathrm{sl}}),
  \qquad
  \mathrm{sinc}(u)=\frac{\sin(\pi u)}{\pi u}.
  \label{eq:slice_collapse_box}
\end{equation}
Substituting \eqref{eq:spatial_factorization} into
\eqref{eq:pdg_signal_spatial} gives
{\setlength{\jot}{1pt}
  \begin{align}
    S^+
    &=
    \sum_v \sum_{\mathbf k,\tau}
    \big(F^+_{\mathbf k,\tau}\big)_v\,
    e^{\,i(\Delta\omega_0)_v\tau}
    e^{-|\tau|/T'_{2,v}}
    \notag\\[-1mm]
    &\quad\times
    \Phi_{v,xy}(\mathbf k_{xy})\,
    \Psi_v(k_z).
    \label{eq:signal_2d_slicecollapse}
\end{align}}

If the phantom is explicitly sampled along the slice direction,
\(\Phi_v(\mathbf k)\) is evaluated by quadrature. For slice samples
\(\{z_{v,q},w_{v,q}\}_{q=1}^{N_q}\), this gives
\begin{equation}
  \Phi_v(\mathbf k)
  \approx
  \Phi_{v,xy}(\mathbf k_{xy})
  \sum_{q=1}^{N_q}
  w_{v,q}\,
  e^{\,i2\pi k_z z_{v,q}} .
  \label{eq:phi_explicit_z}
\end{equation}
Equation~\eqref{eq:phi_explicit_z} is the quadrature version of the same
spatial integral in \eqref{eq:spatial_basis_factor}. When the phantom is not
explicitly resolved along \(z\), the slice-direction integral is evaluated
analytically through \(\Psi_v(k_z)\).

% Equation~\eqref{eq:signal_2d_slicecollapse} preserves the original
% in-plane PDG
% measurement model and extends the slice direction through the factor
% \(\Psi_v(k_z)\). In the present implementation, the analytic collapse provides
% a fast alternative to explicit \(z\) sampling and was found to agree closely
% with explicitly sampled slab simulations in the tested experiments.

\section{Experiments and Results}
All PDG-based simulations were run on a workstation with an NVIDIA RTX 4070 Ti
GPU, an Intel Core i9-13900K CPU, and 64 GB RAM. Proposed simulator
was implemented in
Python/PyTorch with CUDA acceleration; the PDG prepass used a Rust backend.
RF propagator construction,tap extraction, and
voxelwise replay were run on the GPU. Reported runtimes include RF processing,
graph compilation, and signal generation, but exclude image display and file
I/O.

Isochromat Bloch references were run on a Nomad-managed A100 GPU node using a
\(128\times128\) phantom and \(20{,}000\) isochromats per voxel. The largest run
used approximately 73 GB GPU memory, and the FLASH reference required more than
24 h, making higher isochromat counts impractical.

Digital brain phantoms were derived from BrainWeb~\cite{brainweb}. For
species-dependent experiments, water-like and fat compartments were simulated
separately at the same spatial coordinates when both were present, with their
signals summed during signal formation~\cite{Hernando2010ChemShift}. Sequence
parameters are summarized in Table~\ref{tab:sequence_parameters}. Since baseline
PDG accuracy has been established previously~\cite{pdg}, the experiments focus
on the proposed slice-selective RF extension and its effects on RF coupling,
signal formation, species dependence, and runtime.

\subsection{Simulation fidelity and image formation}
\label{sec:results_fidelity}

\begin{figure*}[t]
  \centering
  \includegraphics[width=0.95\textwidth]{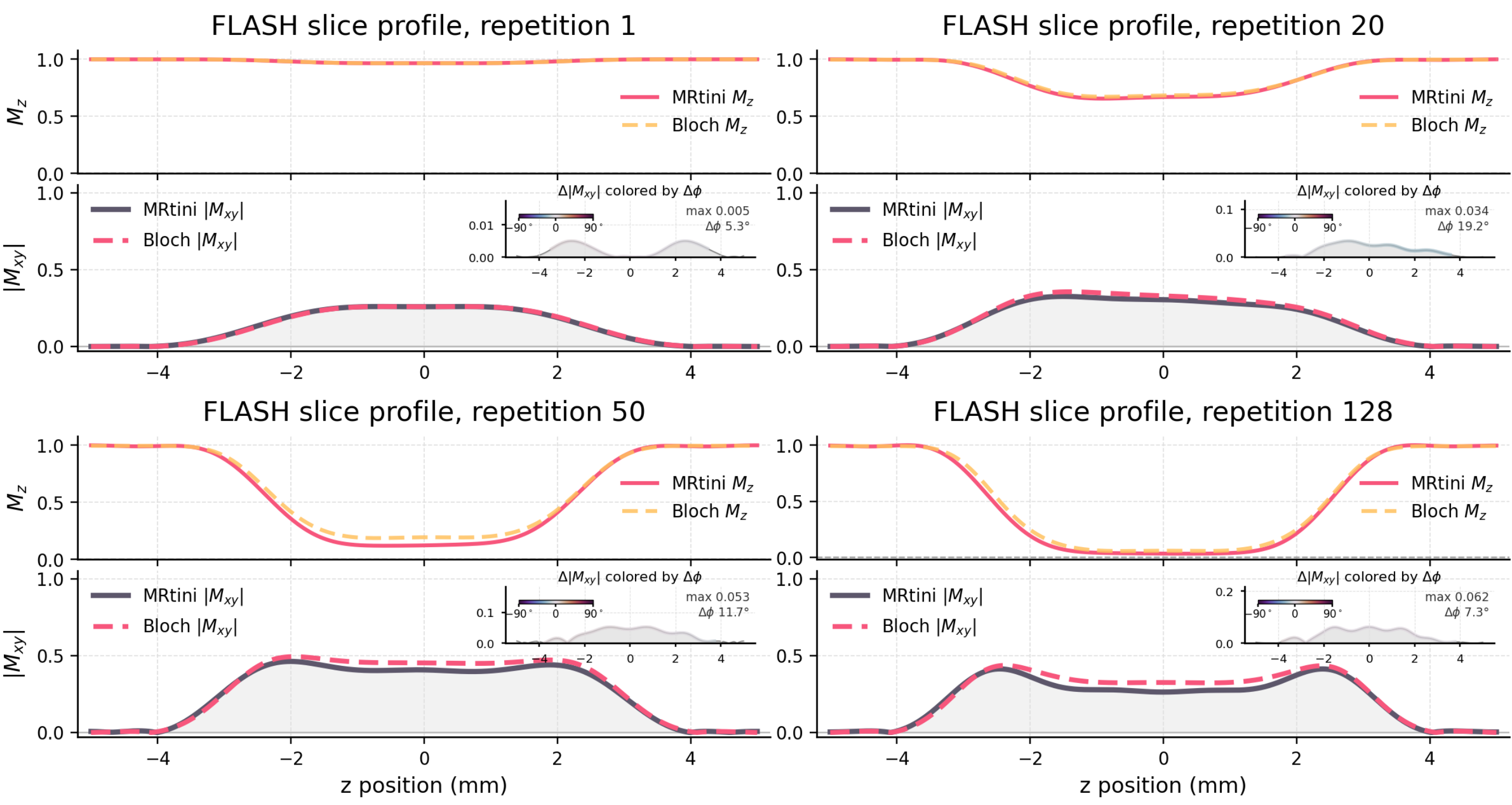}
  \caption{Comparison of repeated-TR slice-profile evolution for a
    FLASH sequence.
    At each repetition, the proposed method uses Bloch-derived RF
    taps inside the
    phase-graph update and reconstructs the resulting magnetization
    profile from the
    retained distributions; the reference is a direct 1-D Bloch
    simulation of the
    same sequence. Panels show $M_z(z)$ and $|M_{xy}(z)|$ at
    repetitions 1, 20, 50,
    and 128. The inset reports the residual error between both methods: the gray
    curve is the pointwise magnitude error in $|M_{xy}|$, and the color overlay
    encodes the local phase difference
    $\Delta\phi=\angle\!\bigl(M_{xy}^{\text{prop}}\overline{M_{xy}^{\text{Bloch}}}\bigr)$.
    The slice domain length $L = 10 mm$, corresponding $d_k=1/L$,
    The proposed model remains close to the Bloch reference until the
  last repetition.}
  \label{fig:results_rf_validation}
\end{figure*}

Figure~\ref{fig:results_rf_validation} compares slice-profile evolution from
the proposed RF-tap/PDG model with a direct 1-D Bloch reference for a FLASH
sequence. Profiles are shown at repetitions 1, 20, 50, and 128 for \(M_z(z)\)
and \(|M_{xy}(z)|\). The proposed model remains close to the Bloch reference
over repeated excitations, with small magnitude and phase residuals. The
comparison used \(N_{\max}=300\) active distributions and RF tap threshold
\(\epsilon=10^{-6}\), matching the settings used in the image simulations.

\begin{figure*}[t]
  \centering
  \includegraphics[width=0.9\textwidth]{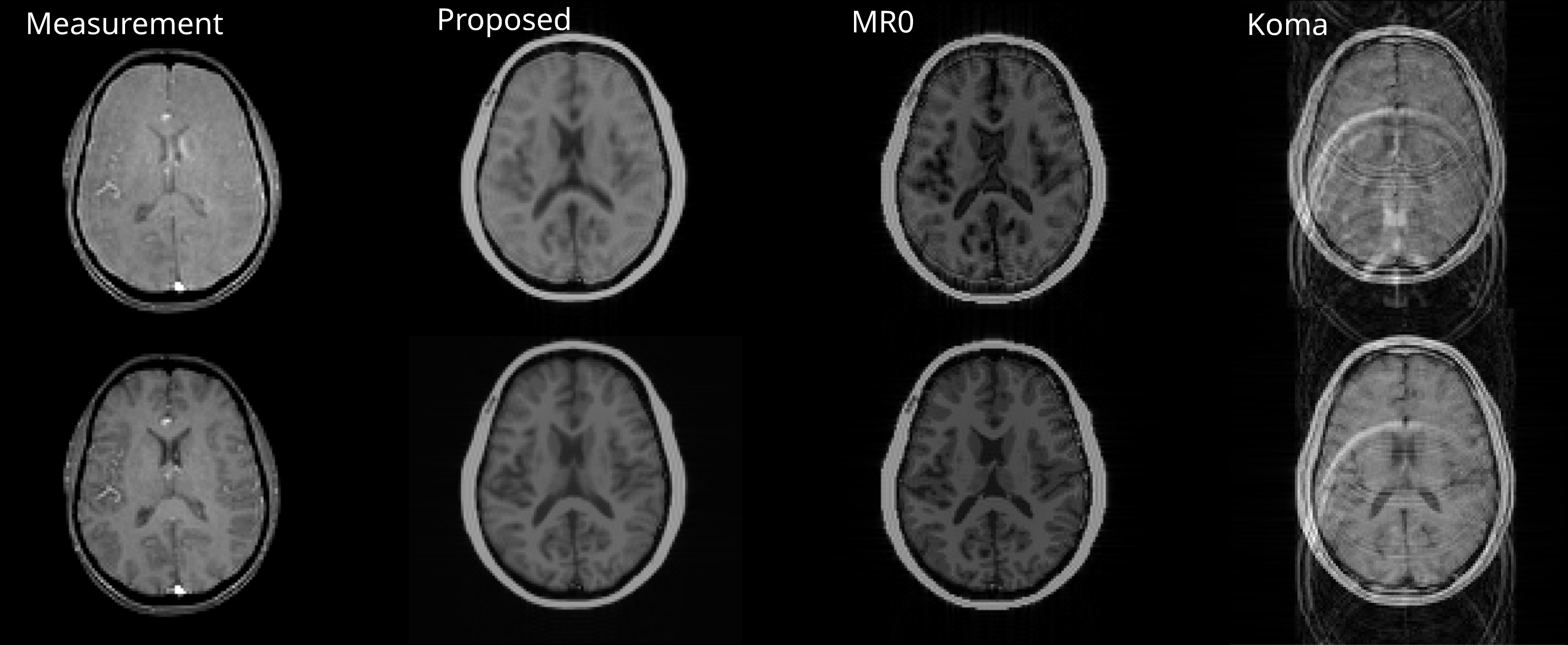}

  \caption{FLASH reconstruction comparison. The top row shows the first
    acquisition repeat, and the bottom row shows the third repeat, where the
    magnetization is closer to steady state. Columns show, from left to right,
    scanner measurement, proposed slice-selective PDG/Bloch model,
    order-local PDG
    baseline (MR0), and full Bloch simulation (Koma). The same FLASH
    sequence was
    used for measurement and simulation
    ($\alpha=15^\circ$, $\mathrm{TR}=10$ ms). The comparison is
    qualitative because
  the measured subject and simulation phantom are not identical.}
  \label{fig:results_flash}
\end{figure*}

\begin{table}[t]
  \centering
  \footnotesize
  \setlength{\tabcolsep}{3pt}
  \caption{Sequence parameters used in the reported simulations and scanner
    comparison with matrix size \(128\times128\),
    \(5\,\mathrm{mm}\)slice thickness,
    \(0.256\times0.256\,\mathrm{m}^2\) FOV, and a maximum tracked
  states capped at\(N_{\max}=300\), unless otherwise specified.}
  \label{tab:sequence_parameters}
  \begin{tabular}{@{}p{0.48\columnwidth}cccc@{}}
    \hline
    Sequence & FA & TR & TE & RO dur. \\
    \hline
    FLASH & \(15^\circ\) & 10 ms & 5 ms & 2 ms \\
    FLASH 1-D profile & \(15^\circ\) & 10 ms &  5 ms & 2 ms \\
    SE-EPI & \(90^\circ/180^\circ\) & 500 ms & 50 ms & 0.5 ms \\
    RARE & \(90^\circ/120^\circ\) & 500 ms & 45 ms & 1 ms \\
    % bSSFP & \(90^\circ\) & 8 ms &4 ms &2 ms \\
    Radial GRE & \(15^\circ\) & 10 ms & 5 ms & 2 ms \\
    Spiral GRE & \(15^\circ\) & 100 ms & 3 ms & 20 ms \\
    \hline
  \end{tabular}
\end{table}
\begin{table}[t]
  \centering
  \caption{runtime comparison for FLASH sequence. The size of the 2D
    phantom used is \(128\times128\)
  and for 3D \(128\times128\times 32\)}
  \label{tab:runtime}
  \setlength{\tabcolsep}{4pt}
  \begin{tabular}{@{}p{0.30\columnwidth}cccc@{}}
    \hline
    Method & RF kernel & Prepass & Replay & Total \\
    \hline
    Order-local PDG & -- & 0.131 & 6.713 & 6.844 s \\
    Proposed (Order-local + off-res) & 0.6 & 0.699 & 4.283 & 5.582 s \\
    Proposed method & 0.599 & 1.017 & 23.246 & 24.862 s \\
    Proposed method(3D phantom) & 0.31 & 0.12 & 63.3 & 63.73 s\\
    Koma(Bloch baseline) & -- & -- & -- & 180.68 s\\
    \hline
  \end{tabular}
\end{table}
The effect of the RF-resolved model on reconstructed images is shown in
Fig.~\ref{fig:results_flash}, which compares FLASH reconstructions from an
in-vivo scanner measurement and three simulation pipelines using gammaSTAR
sequence definition, with the measurement acquired on a Siemens
VIDA-fit 3T system. Because the measured
subject and the digital phantom are not identical, the comparison is qualitative
rather than voxelwise. The proposed slice-selective PDG/Bloch model, the
scanner measurement, and the Bloch reference show a similar lighter contrast
appearance, whereas the order-local PDG baseline appears slightly darker in this
example.

\begin{figure}[t]
  \centering
  \includegraphics[width=0.9\columnwidth]{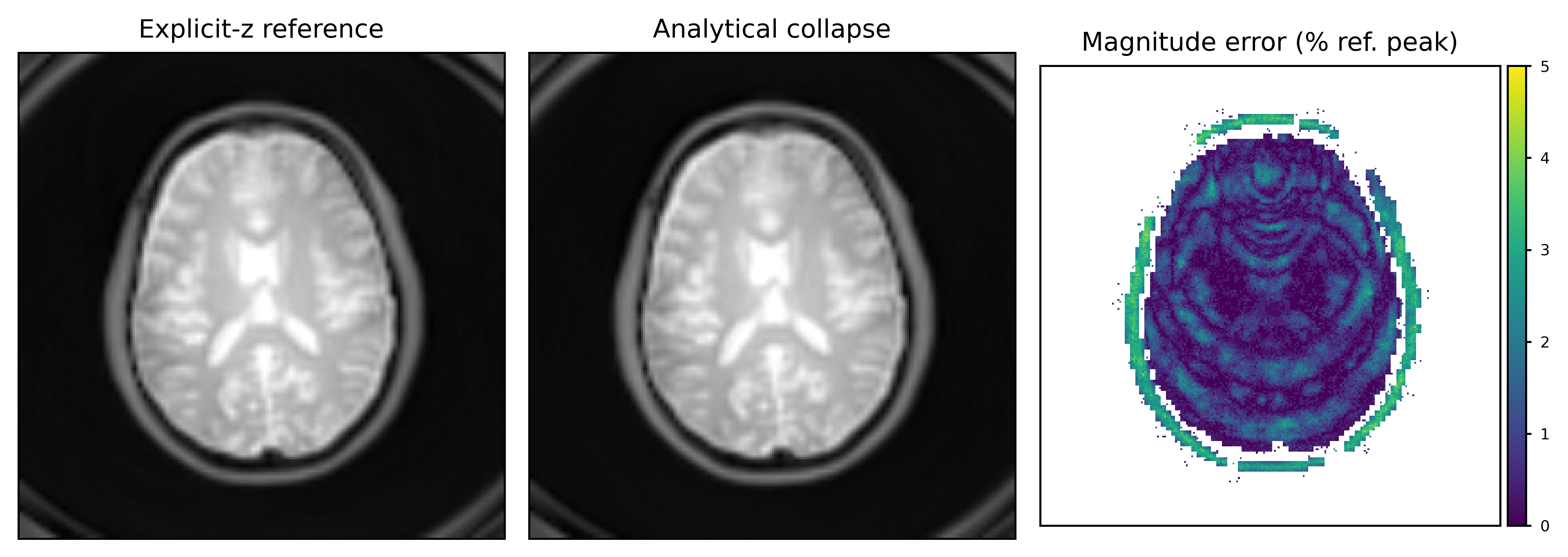}
  \caption{Comparison of explicit-$z$ sampling and analytic slice
    collapse. (Spiral Sequence)
  with SSIM of 99.5\% .}
  \label{fig:results_slice_collapse}
\end{figure}

The slice-aware measurement model of Sec.~\ref{sec:measurement_model} was
examined in both explicit-$z$ and analytic-collapse form.
Figure~\ref{fig:results_slice_collapse} compares these two realizations for
a Spiral sequence. The resulting reconstructions are very similar, with
small residual differences. The measured SSIM was 99.5\%.

\begin{figure}[t]
  \centering
  \includegraphics[width=\columnwidth]{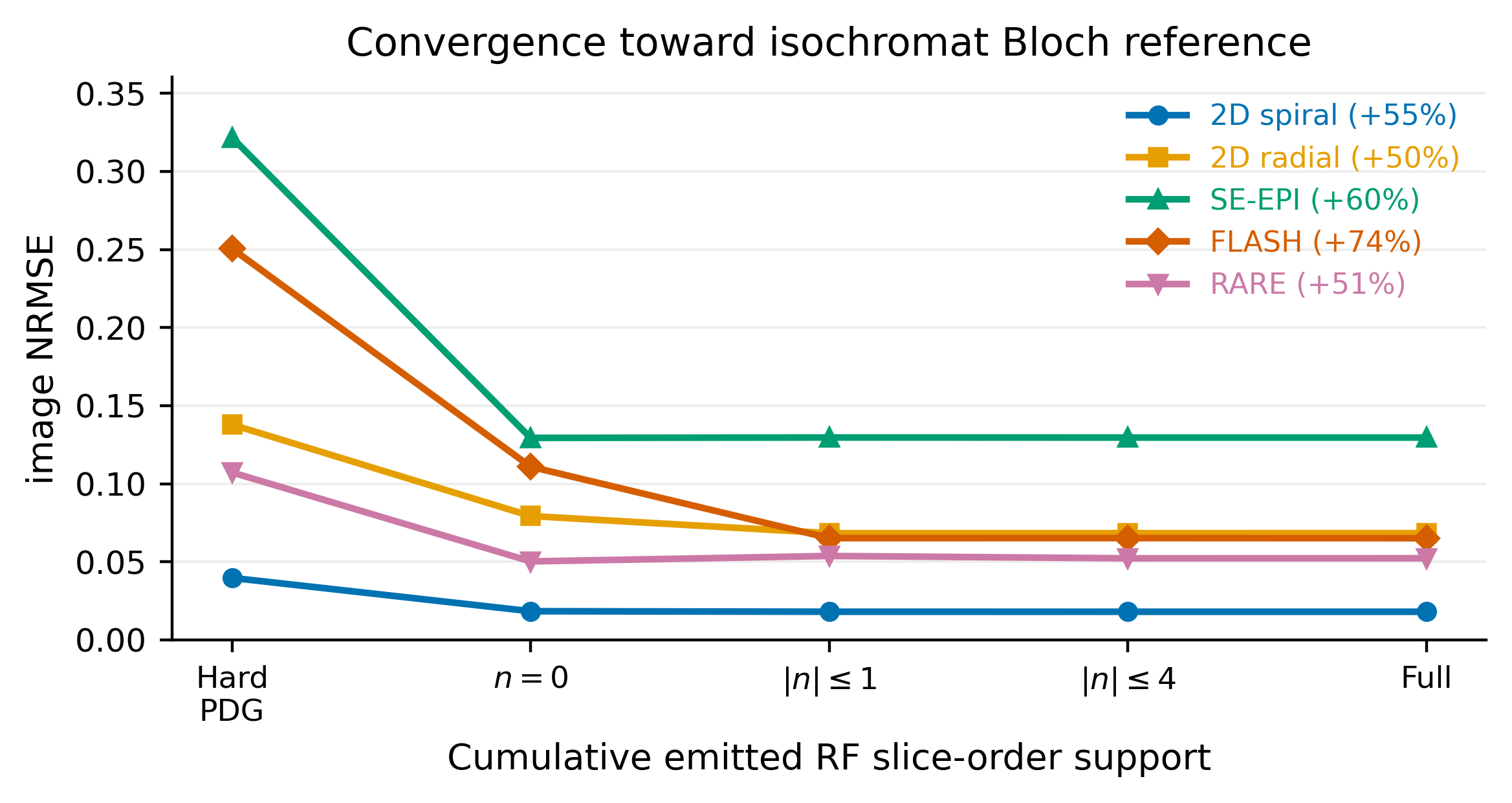}
  \caption{NRMSE was computed between the reconstructed images and the
    corresponding isochromat Bloch reference after
    gammaSTAR reconstruction. The compared models are the hard-pulse
    PDG baseline, the center RF slice-order contribution $((n=0))$,
    cumulative emitted-signal at different orders including $(|n|\leq1)$ and
    $(|n|\leq4)$, and the full slice-aware PDG signal. The legend labels
    shows the relative NRMSE improvement with respect to the hard-pulse
  baseline for each sequence.}
  \label{fig:kz_order_convergence}
\end{figure}

\begin{figure*}[t]
  \centering
  \includegraphics[width=0.92\textwidth]{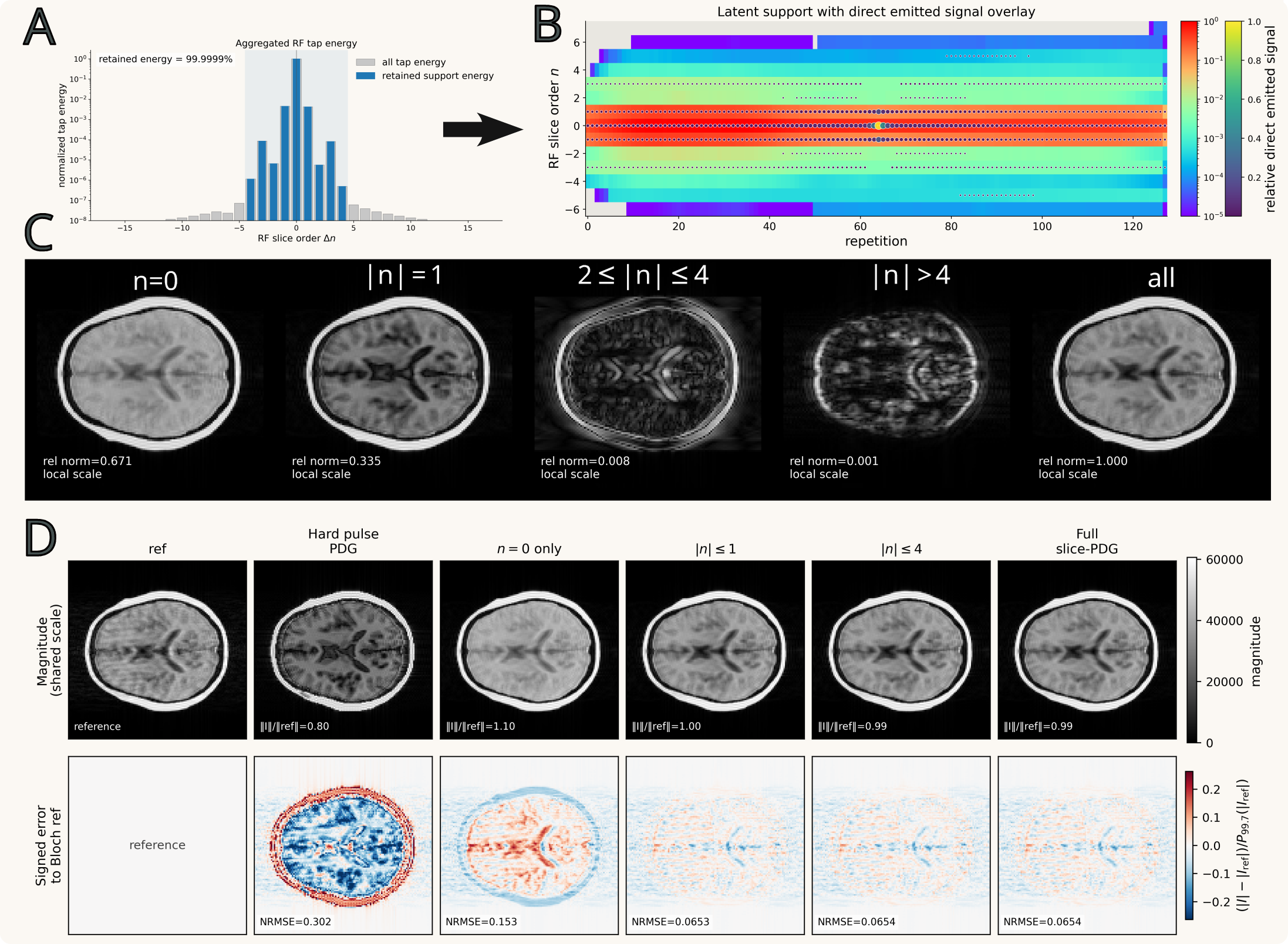}
\caption{RF slice-order analysis and FLASH image validation.
(A) RF tap energy versus coupling increment \(\Delta n\), with retained taps in
blue. (B) Slice-order latent signal map with ADC-emitting states marked.
(C) Emitted order-band reconstructions and their relative norm with respect to
the full slice-aware PDG image. (D) Comparison with the isochromat Bloch
reference; bottom row shows signed magnitude error.}
    % The hard-pulse PDG
    % baseline shows larger contrast and spatial errors, while including
    % the dominant RF slice-order contributions substantially reduces the
    % discrepancy to the Bloch reference. Even at 20k isochromats per voxel,
    % the reference is still a numerical quadrature rather than an
    % analytic ground truth, and
    % further refinement was computationally infeasible.
  \label{fig:analysis_flash}
\end{figure*}

% In order to verify that the nonzero RF orders indeed contribute to
% the measured image
% The RF slice-orders(Fig.~\ref{fig:analysis_flash}) were analyzed. The
% aggregated tap spectrum shows the local RF coupling increments
% $(\Delta n)$ retained
% during graph construction, while the latent activity map in
% Fig.~\ref{fig:analysis_flash}-B identifies slice-order
% states that remain relevant either through direct ADC emission or through
% later mixing into emitting states(Latent Signal). States that
% directly contribute to the signal are overlaid on the same map as dots.
% % Thus, orders with weak direct emission can still
% % affect the measured signal by rephasing or merging into lower-order
% % states at later repetitions. This explains why the $(n=0)$ contribution
% % of the slice-aware model is not equivalent to the hard-pulse $(n=0)$
% % approximation: it can contain pathways that previously visited
% % nonzero slice orders before returning to an emitting state.
% For other sequences, image-level effect were quantified by
% progressively adding emitted order bands and comparing the resulting
% reconstructions with isochromat Bloch references
% (Fig.~\ref{fig}). This experiment demonstrates that the dominant
% improvement over the hard-pulse PDG baseline arises from the
% Bloch-resolved center and first slice-order contributions, with
% higher orders producing smaller sequence-dependent corrections.
To assess whether the RF-induced nonzero slice orders contribute to the measured
image, the retained slice-order structure was analyzed
(Fig.~\ref{fig:analysis_flash}). The aggregated tap spectrum in (A)
shows the local RF
order increments \(\Delta n\) retained during graph construction. The latent
activity map in (B) identifies slice-order states that either emit
signal directly at
ADC events or remain relevant through later mixing into emitting
states. Directly
emitting states are overlaid as dots. The image-level contribution
(C) of slice orders was
quantified by progressively adding emitted order bands and comparing the
resulting reconstructions with isochromat Bloch references (D).
Results from additional sequences are summarized in Fig.
~\ref{fig:kz_order_convergence}. The results show that most of the
improvement over the hard-pulse PDG baseline comes from the Bloch-resolved
center and first slice-order bands, while higher orders provide smaller,
sequence-dependent corrections.

\begin{figure}[t]
  \centering
  \includegraphics[width=\columnwidth]{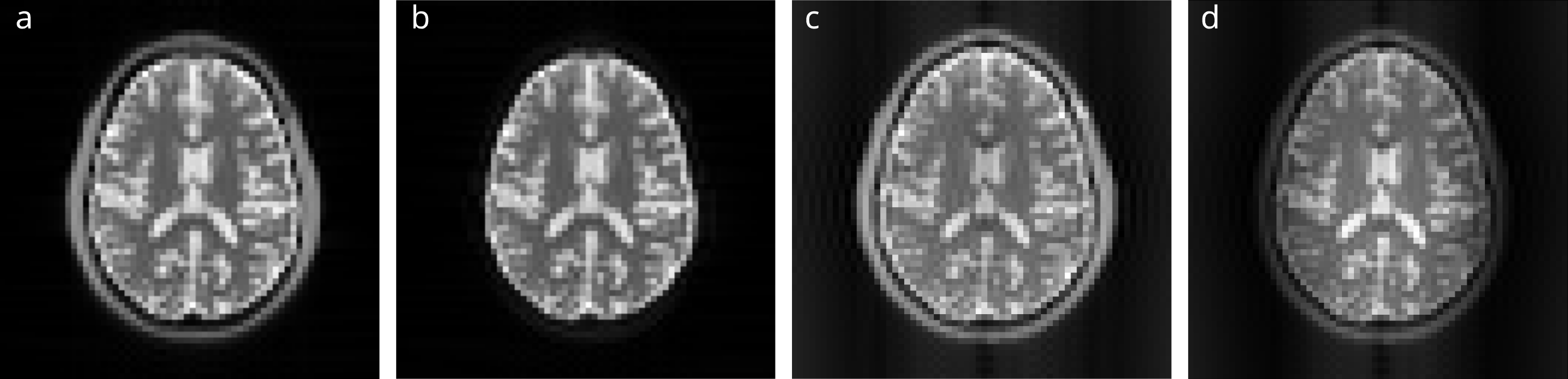}
  \caption{Fat-suppressed SE-EPI simulation.
    (a,b) Proposed RF-resolved model without and with fat-suppression
    preparation.
    (c,d) Order-local PDG baseline without and with the same preparation.
    The RF-resolved model includes species-dependent frequency offsets during RF
    propagation, whereas the order-local baseline does not; the latter therefore
    shows a stronger contrast change after the nominal fat-suppression pulse and
  incomplete suppression of the fat component.}
  \label{fig:fatsat_epi}
\end{figure}

\subsection{Slice-direction structure, system effects, and runtime}
\label{sec:results_extensions}

\begin{figure}[t]
  \centering
  \includegraphics[width=\columnwidth]{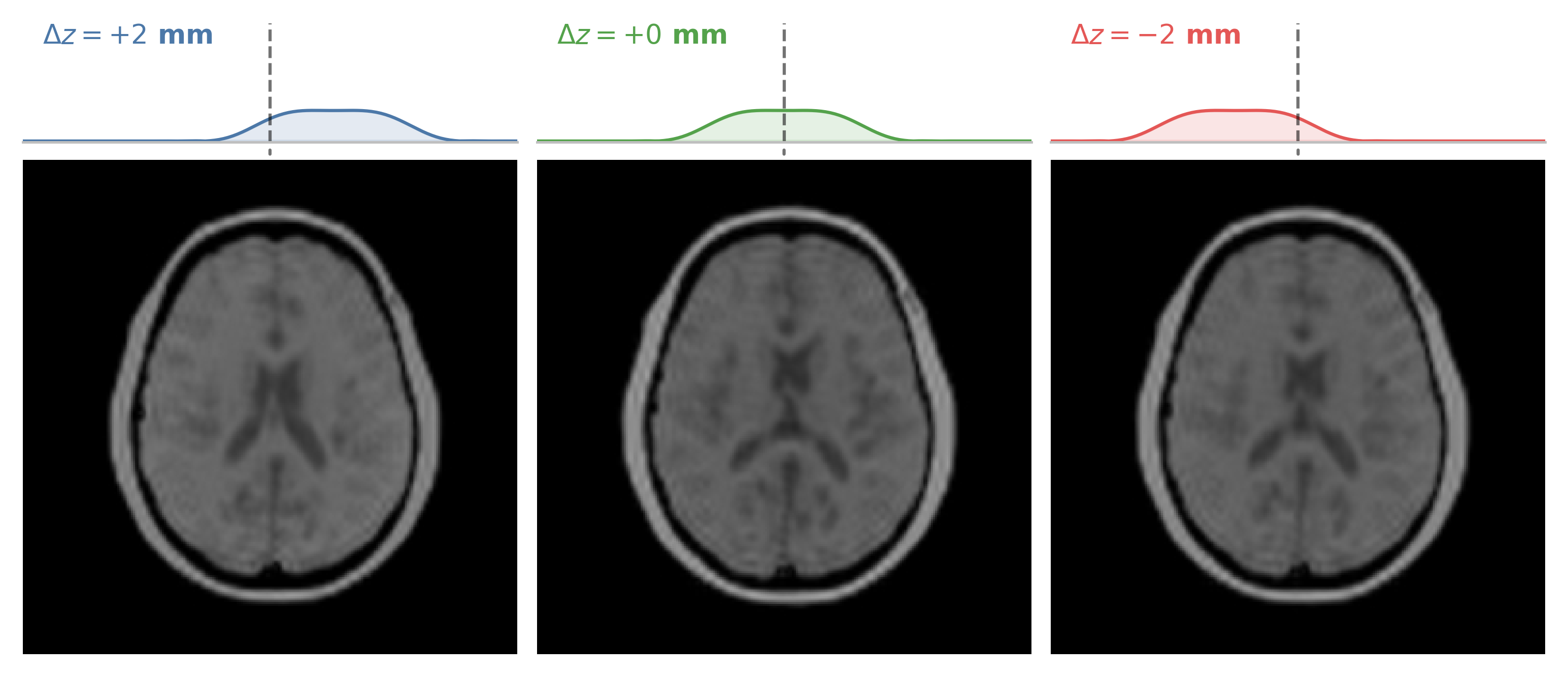}
  \caption{Through-slice mixing under explicit-$z$ sampling.
    The change in reconstructed anatomy reflects the changed overlap between the
    slice profile and the heterogeneous object along $z$. This serves as an
    important check that the explicit-$z$ signal model responds correctly to
  slice-position changes.}
  \label{fig:results_partial_volume}
\end{figure}

To examine the use of explicit sampling along the slice direction, a 3D
phantom was simulated over an FOV larger than the excited slab, using 32
samples per voxel along $z$.
The slice profile was shifted by \(\pm 2\) mm. The resulting
reconstructions changed visibly with slice position as illustrated in
Figure~\ref{fig:results_partial_volume}. Additionally the framework in
not limited to 2D sequences and supports full 3D gammaSTAR sequences
with volumetric phantoms.

Figure~\ref{fig:fatsat_epi} compares SE-EPI simulations with and without a
fat-suppression preparation. Panels (a,b) show the proposed RF-resolved model,
where water and fat are simulated with species-dependent frequency offsets
during the RF spans. The fat-suppression pulse reduces the fat component while
leaving the water-dominated contrast largely consistent with the corresponding
non-suppressed case. WHile order-local PDG baseline (Panels c and d)
under the same sequence and
state-pruning settings without off-resonance modeling, the
preparation changes the
overall contrast more visibly and the fat component is not fully suppressed.

Table~\ref{tab:runtime} summarizes average runtime over 10 runs.
The proposed slice-selective model is more expensive than the order-local
baseline, as expected, but remains cheaper than full Bloch simulation in the
tested settings.

% \begin{figure}[t]
%   \centering
%   \includegraphics[width=\columnwidth]{figs/tap_energy_spectrum_comparison.png}
%   \caption{Comparison using measured 3D and reduced 2D $B_0$ maps.
%   Placeholder panels}
%   \label{fig:results_b0_3d}
% \end{figure}

\section{ Discussion and Conclusion}

A central aspect of the present work is the establishment of a common
formalism between
two MRI simulation viewpoints that are developed in parallel rather than
together: PDG, as a spatially resolved extension of EPG, and
Bloch-resolved RF modeling
for shaped slice-selective and off-resonant pulse dynamics.
The value of the method is therefore not only improved slice-profile
modeling, but the ability to combine RF fidelity, pathway history, and spatially
resolved image formation in one sequence-consistent simulator.
This unification matters most in regimes where RF physics, spatial encoding,
object heterogeneity, and echo-pathway history interact. Examples include
slice-profile-dependent contrast, fat suppression, chemical-shift-sensitive
excitation, refocusing trains, and simulations driven by measured field
inhomogeneity. Because the simulator retains the PDG state graph, the resulting
signal can also be analyzed in terms of contributing echo pathways rather than
only as a final image or time series. This provides a way to study how
Bloch-resolved RF effects redistribute magnetization across pathways and how
those pathways later rephase at readout. The same framework can
operate in full 3D with
3D phantoms and sequences
% , including maps acquired on the Delori system,
enabling more system-specific simulation than
slice-agnostic RF models.

The 1-D FLASH slice-profile experiment provides a useful stress test for this
coupling. A single slice-selective RF pulse can populate many nonzero
$k_z$-orders, so it was not obvious a priori that a truncated PDG graph would
preserve the accumulated slice profile over many repetitions. In the tested
sequence, the proposed RF-tap/PDG representation remained close to the
direct Bloch reference through repetition 128 while retaining a few hundred
active distributions with the same reduced RF-tap support and state cap
used later for image formation. This suggests that the
slice-profile dynamics remain sufficiently compressible for PDG-style graph
propagation.

In the FLASH image reconsturction experiment, the proposed model
follows the measured image
and the Bloch reference more
closely in overall contrast impression, which is consistent with the inclusion
of slice-profile effects in the RF model.The Koma result provides a
full Bloch reference under the same nominal
sequence, but the short-TR RF-spoiled FLASH setting is challenging for finite
isochromat sampling despite using the highest available resolution
for the digital phantom.
% The short-TR RF-spoiled FLASH case also illustrates a practical limitation of
% finite-isocromat Bloch simulation.For Koma, the first three repeats
% showed stronger
% transient artifacts, and additional repeats did not improve the reconstruction
% appreciably. Increasing the TR to 100 ms reduced these artifacts, but also
% changed the steady-state contrast and therefore was not comparable to the
% measured $\mathrm{TR}=10$ ms experiment.
The observed artifacts are consistent
with insufficient resolution of spoiler-induced intravoxel dephading: residual
transverse magnetization can persist numerically across repetitions and leak
into the signal. This is a typical resolution limitation of finite-spin Bloch
simulation in this setting.
PDG-based methods are robust in this regard because dephasing and
spoiling are represented through tracked phase-distribution states rather than
only through explicit dense spin/isochromat sampling.

The explicit-$z$ formulation is relevant when through-slice structure
is of interest.
The slice-shift experiment demonstrates the role of explicit \(z\)-sampling:
the reconstructed signal changes with the physical overlap between the
slice-selective profile and the heterogeneous object. This behavior would not be
captured by a slice-agnostic hard-pulse model, Additionally, this
provides a practical check that the
slice-profile RF coupling and the explicit-$z$ measurement model
are acting consistently inside the PDG framework.
% Unlike the analytic collapse,
% which assumes a prescribed through-slice support, explicit $z$ sampling can
% represent heterogeneous structure along the slice direction directly.

% . Because off-resonance and
% chemical shift are included during the RF-span solve, water and fat experience
% different effective RF propagators. In the order-local baseline, the RF step
% does not distinguish these species during the pulse, so the nominal
% fat-suppression module also perturbs water-like magnetization and leaves
% residual fat signal.
The fat-suppression experiment illustrates why species-dependent RF response is
not a separate add-on in the proposed formulation.
Recent work has extended PDG-style simulation to
off-resonant RF pulses by using an effective-axis rotation model for
block pulses
\cite{Dietz2025OffResPDG}. This enables applications such as fat saturation,
binomial water excitation, and WASABI-style preparations, the formulation
assumes constant RF amplitude and retains the instantaneous-pulse treatment.
In present work, species-dependent off-resonance
enters through the same Bloch-resolved RF propagator used for
slice-profile modeling, rather than as a separate correction.

The proposed slice-aware PDG simulation consistently reduced the
discrepancy to the isochromat Bloch reference compared with the
order-local hard-pulse PDG baseline. The order-band analysis further
shows that most of the improvement over the hard-pulse PDG baseline
comes from the Bloch-resolved center and first slice-order bands,
consistent with the concentration of RF tap energy in the lowest
orders. These pathways can contribute directly to the signal or
rephase into emitting states during later repetitions. Higher orders
provide smaller, sequence-dependent corrections, and the error
generally plateaus after the first few bands. This again supports the
interpretation that the relevant slice-profile dynamics are sparse in
RF-order space.

In the order-local setting, the cached-tap implementation is slightly faster
than the original PDG baseline, while also allowing off-resonance and
species-dependent RF responses. For
the slice-selective model, the main remaining cost is the larger compiled graph induced by nonzero slice-order offsets. 
During replay, each retained distribution is evaluated by accumulating contributions
from its ancestor edges; slice-selective RF increases this edge count through
additional RF-spawn branches. Replay can therefore become dominated by edge
traversal and accumulation. More efficient compiled-graph execution, for
future work.

\paragraph{Generality and limitations.}
The formulation is not limited to isolated slice-selective
excitations: interleaved multi-slice acquisitions can be handled by
solving and caching slice-specific RF spans, with additional reuse
possible when slices differ only by spatial shifts of the same
profile. SMS and adiabatic pulses are also compatible in principle
because the RF-span solver operates on the full complex RF and
gradient raster; however, they may require broader tap support,
denser $(B_0/B_1)$ binning, and larger graph/state budgets. Long
adiabatic preparation or inversion pulses may also require relaxation
during RF, which is neglected here. Full SMS image simulation would
further require receive-coil and slice-unaliasing models, which were
not evaluated. 

The in-vivo FLASH comparison in Fig.~\ref{fig:results_flash} is also affected
by physiological effects not yet modeled. In
particular, blood inflow after excitation can change the measured contrast.
Motion and flow could be incorporated through
time-resolved PDG state evolution, but explicit 3-D transport of magnetization
is not yet implemented. Finally, although the PDG machinery is compatible
with automatic differentiation, end-to-end differentiable sequence
optimization remains future work.

In short, the proposed method introduces slice profiles directly into the phase-distributions by converting
waveform-resolved RF spans into sparse phase-graph coupling kernels, the method
retains the graph-based efficiency and pathway structure of PDG while capturing
RF effects that are absent in order-local hard-pulse models,
remaining substantially compact compared to full isochromat Bloch simulation. This provides a practical
route toward sequence-consistent simulation in areas where shaped RF pulses,
off-resonance, chemical shift, and through-slice structure jointly affect the
measured signal.

% ==============================================================
%=
%= New Subsection slice-selective RF induces cross-order coupling
%=
%===============================================================

%% add a page break before the appendices
% \clearpage

\end{document}